\documentclass[aps, pra, amsmath, amssymb, showpacs, superscriptaddress, reprint]{revtex4-1}

\usepackage{graphicx}
\usepackage{natbib}

\begin{document}

\title{AC Stark Shifts of Dark Resonances Probed with Ramsey Spectroscopy }

\author{J. W. Pollock}
\email{jyp@nist.gov}
\affiliation{National Institute of Standards and Technology, Boulder, Colorado 80305, USA}
\affiliation{University of Colorado, Boulder, Colorado 80309-0440, USA}

\author{V. I. Yudin}
\affiliation{Novosibirsk State University, ul. Pirogova 2, Novosibirsk, 630090, Russia }
\affiliation{Institute of Laser Physics SB RAS, pr. Akademika Lavrent'eva 13/3, Novosibirsk, 630090, Russia }
\affiliation{Novosibirsk State Technical University, pr. Karla Marksa 20, Novosibirsk, 630073, Russia }

\author{M. Shuker}
\affiliation{National Institute of Standards and Technology, Boulder, Colorado 80305, USA}
\affiliation{University of Colorado, Boulder, Colorado 80309-0440, USA}

\author{M. Yu. Basalaev}
\affiliation{Novosibirsk State University, ul. Pirogova 2, Novosibirsk, 630090, Russia }
\affiliation{Institute of Laser Physics SB RAS, pr. Akademika Lavrent'eva 13/3, Novosibirsk, 630090, Russia }
\affiliation{Novosibirsk State Technical University, pr. Karla Marksa 20, Novosibirsk, 630073, Russia }

\author{A. V. Taichenachev}
\affiliation{Novosibirsk State University, ul. Pirogova 2, Novosibirsk, 630090, Russia }
\affiliation{Institute of Laser Physics SB RAS, pr. Akademika Lavrent'eva 13/3, Novosibirsk, 630090, Russia }

\author{X. Liu}
\affiliation{National Institute of Standards and Technology, Boulder, Colorado 80305, USA}

\author{J. Kitching}
\affiliation{National Institute of Standards and Technology, Boulder, Colorado 80305, USA}
\affiliation{University of Colorado, Boulder, Colorado 80309-0440, USA}

\author{E. A. Donley}
\affiliation{National Institute of Standards and Technology, Boulder, Colorado 80305, USA}
\affiliation{University of Colorado, Boulder, Colorado 80309-0440, USA}

\date{\today}

\begin{abstract}
The off-resonant AC Stark shift for coherent population trapping (CPT) resonances probed with Ramsey spectroscopy is investigated experimentally and theoretically. Measurements with laser-cooled $^{87}$Rb atoms show excellent quantitative agreement with a simple theory. The shift depends on the relative intensity of the two CPT light fields, but depends only weakly on the total intensity. Since the origin of the shift is through couplings of the interrogation light to off-resonant excited state hyperfine levels, the size and sign of the shift depend on the specific interrogation scheme. The theory also shows that for several commonly used interrogation schemes it is possible to minimize the off-resonant light shift or its dependence on the CPT intensity ratio by properly selecting the system parameters. 
\end{abstract}


\maketitle

\section{Introduction}
\label{Introduction}

The AC Stark shift, or light shift, is a light-induced change in the energy level structure of atoms with important implications to atomic timing. While this shift plays an important role in continuously pumped vapor-cell atomic clocks, it is not traditionally present in high-accuracy beam and fountain clocks where the coherent manipulation of the atoms is done with microwave fields, and optical fields are only present during state preparation and readout. However, AC Stark Shifts play an important role in contemporary optical clocks \cite{Ludlow_2015} and microwave clocks based on coherent population trapping (CPT) \cite{Alzetta_1976,arimondo_1996,Vanier_2005,Shah_2010} in which light fields modulated at microwave frequencies induce the coherent excitation of the atoms and can contribute significant light shifts.

Light shifts \cite{Arditi_PhysRev_1961} in continuously-probed (CW) vapor-cell CPT clocks have been studied in detail \citep{Vanier_1998,Wynands1999,vanier_1999,levi_2000,Zhu_PTTI_2000,Miletic_APB_2012,Hafiz_2016,Yun_2017}, where they cause frequency biases sensitive to the optical detuning, total CPT intensity, cell temperature \cite{Miletic_APB_2012}, as well as the intensity ratio between the two components of the CPT light field \cite{vanier_1999,levi_2000,Zhu_PTTI_2000, Miletic_APB_2012, Yun_2017} (Fig. \ref{image:Lambda-scheme}.a). Operating at an optimized intensity ratio can reduce the frequency dependence on intensity \cite{Zhu_PTTI_2000, Zibrov_2010, Miletic_APB_2012}, however residual shifts remain \cite{Yun_2017}. 

CPT resonances can also be probed with Ramsey spectroscopy \cite{Thomas_PRL_1982, Hemmer_JOSAB_1989, Zanon_EFTF2004, Zanon_PRL_2005, Guerandel_IEEETM_2007, Castagna_2009,Kozlova_IEEE_2014,Yano_PRA_2014,Yano_APExp_2015}, with which the atoms are probed by two pulses of length $\tau_1$ and $\tau_2$ separated by a dark interval of length $T$ (Fig. \ref{image:Lambda-scheme}.b). In Ramsey spectroscopy, the light shift can be divided into a resonant and off-resonant component. The resonant shift involves the resonant interaction of the two light fields and three atomic levels (Fig. \ref{image:Lambda-scheme}.a). Previous work \cite{Hemmer_JOSAB_1989,Shahriar_1997,Pati_2015} has shown that resonant shifts result from incomplete dark-state formation during the first Ramsey pulse, and vanish when the product of the total intensity and the first Ramsey pulse duration are sufficiently large \cite{Liu_APL_2017}. This leaves the off-resonant light shift as the dominant systematic shift. This shift involves the interaction of all optical field components with all detuned atomic energy levels that couple with the light (normally the ground- and excited-state hyperfine structure).

Ramsey spectroscopy has a few general advantages over CW interrogation for CPT clocks, but the trade-offs between the two excitation protocols are subtle and require careful analysis. Several Ramsey CPT studies have shown reduced light shifts compared to CW interrogation at the same intensity \cite{Castagna_2009,Yano_PRA_2014,Hafiz_2017}. However, in CPT clocks based on continuous excitation, the light illuminates the atoms for a longer duration, and therefore a considerably lower intensity is needed to achieve the same optical pumping efficiency. Thus, both CW and Ramsey schemes result in a comparable absolute shift at optimal operating conditions. 

However, in Ramsey CPT clocks the sensitivity of the light shift to intensity variations is significantly reduced when compared to CW experiments \cite{Yano_PRA_2014,Yano_APExp_2015,Liu_PRApplied_2017}. As we show in this work, the off-resonant shift depends weakly on intensity. Thus, in Ramsey CPT clocks, light shifts can have a lower sensitivity to intensity variations even if the magnitude of the shift is comparable to low-intensity continuous excitation. Additionally, in Ramsey spectroscopy various schemes can be implemented to eliminate light shifts by using pulses with tailored phases and frequencies \cite{Yudin_2010,yudin_2017_GABRS,sanner_2018_autobalance}.

Cold-atom CPT clocks \cite{Esnault_PRA_2013, blanshan2015light, Liu_APL_2017,Liu_PRApplied_2017} offer a clean system for studying these light shifts since buffer-gas shifts \cite{Arditi_PhysRev_1958,Liu_PRA_2013} are avoided, narrow resonances with long coherence periods are observed \cite{Liu_APL_2017,Liu_PRApplied_2017}, and nearly complete dark states can be formed.  By mitigating known shifts, the cold-atom CPT clock provides an excellent platform for performing precise comparisons between theory and experiment as well as exploring performance limits. In this work, we use a cold-atom CPT clock to experimentally investigate a simple theory for the off-resonant shifts. Comparisons between experimental measurements and theoretical models are in excellent quantitative agreement. We show that the interrogation can be tuned to minimize the off-resonant shifts or reduce their sensitivity to various experimental parameters. These results should allow for improved performance across a broad range of future CPT atomic clocks. 

\section{Theory}
\label{Theory}

\begin{figure}[t]
    \includegraphics[width=8.6cm]{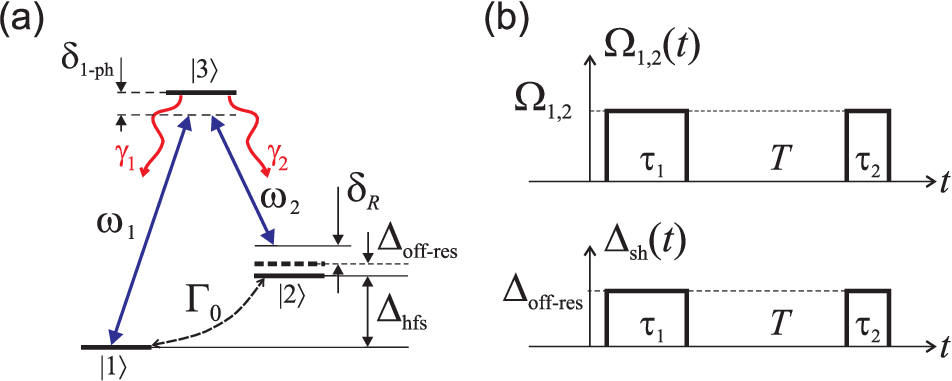}
    \caption{
    \label{image:Lambda-scheme}
    (a) Atomic three-level $\Lambda$ system. (b) Schematic time dependencies in the Ramsey scheme of the Rabi frequencies $\Omega_{1, 2}$ and the frequency shifts $\Delta_{\text{off-res}}$. The pulse lengths are typically: $\tau_1 = 3$ ms, $T = 4 - 16$ ms, and $\tau_2 = 50\ \mu$s.
    }
\end{figure}

As a model, we consider CPT resonances formed in a three-level $\Lambda$ system under interaction with two monochromatic fields
\begin{equation}\label{bichromatic field}
E(t) = E_1 \cos(\omega_1 t) + E_2 \cos(\omega_2 t)
\end{equation}
each detuned by $\delta_{\text{1-ph}}$ from a short-lived excited electronic state $|3\rangle$ with a spontaneous decay rate $\gamma$. The CPT resonance forms when the difference between the optical frequencies ($\omega_{1}-\omega_{2}$) is near the microwave-frequency transition between the long lived lower energy levels $|1\rangle$ and $|2\rangle$ or $\omega_{2}-\omega_{1}\approx\Delta_{\text{hfs}}$ (Fig.~\ref{image:Lambda-scheme}.a). The dynamics of the $\Lambda$ system in the rotating wave approximation are described by the differential equation system for the density matrix components:
\begin{align}\label{rho_eqs}
&[\partial_t+\gamma_\text{opt}-i\delta_\text{1-ph}]\rho_{31}=\frac{i\Omega_1}{2}(\rho_{11}-\rho_{33})+\frac{i\Omega_2\rho_{21}}{2}\nonumber \\
&[\partial_t+\gamma_\text{opt}-i\delta_\text{1-ph}]\rho_{32}=\frac{i\Omega_2}{2}(\rho_{22}-\rho_{33})+\frac{i\Omega_1\rho_{12}}{2}\nonumber\\
&[\partial_t+\Gamma_0 -i\delta_R]\rho_{12}= \frac{i}{2}( \Omega^{\ast}_1\rho_{32}-\rho_{13}\Omega^{}_{2} ) \\
&[\partial_t+\Gamma_0 ] \rho_{11}=\gamma_1\rho_{33}+\frac{\Gamma_0}{2} \text{Tr}\{\hat{\rho}\}+\frac{i}{2}(\Omega^{\ast}_1\rho_{31}-\rho_{13}\Omega^{}_{1})\nonumber\\
&[\partial_t+\Gamma_0 ] \rho_{22}=\gamma_2\rho_{33}+\frac{\Gamma_0 }{2}\text{Tr}\{\hat{\rho}\}+\frac{i}{2}(\Omega^{\ast}_2\rho_{32}-\rho_{23}\Omega^{}_{2})\nonumber\\
&[\partial_t+\Gamma_0 +\gamma] \rho_{33}= \frac{i}{2}(\Omega_1\rho_{13}-\rho_{31}\Omega^{\ast}_{1})+\frac{i}{2}(\Omega_2\rho_{23}-\rho_{32}\Omega^{\ast}_{2})\nonumber,
\end{align}
with the conditions $\rho^{}_{jk}=\rho^{\ast}_{kj}\;(j,k=1,2,3)$ and $\text{Tr}\{\hat{\rho}\}=\rho_{11}+\rho_{22}+\rho_{33}=1.\nonumber$

Here $\Omega_{1}(t)=d_{31}E_{1}(t)/\hbar$ and $\Omega_{2}(t)=d_{32}E_{2}(t)/\hbar$ are the Rabi frequencies for the transitions $|1\rangle\leftrightarrow|3\rangle$ and $|2\rangle\leftrightarrow|3\rangle$, respectively ($d_{31}$ and $d_{32}$ are the reduced dipole matrix elements for these transitions); $\gamma_\text{opt}$ is rate of decoherence of the optical transitions $|1\rangle\leftrightarrow|3\rangle$ and $|2\rangle \leftrightarrow|3\rangle$ ($\gamma_\text{opt}=\gamma/2$ for pure spontaneous relaxation); $\gamma_1$ and $\gamma_2$ are spontaneous decay rates ($\gamma_1+\gamma_2=\gamma$ for a closed $\Lambda$ system); $\Gamma_0$ is the slow ($\Gamma_0\ll \gamma,\gamma_\text{opt}$) rate of relaxation to the equilibrium isotropic ground state; $\delta_R = \omega_{2}-\omega_{1}-\Delta_{\text{hfs}}-\Delta_{\text{off-res}}(t)$ is two-photon (Raman) detuning, where $\Delta_\text{off-res}(t)$ is an additional shift between levels $|1\rangle$ and $|2\rangle$ present only when the CPT light is on (Fig. \ref{image:Lambda-scheme}.b). 

$\Delta_{\text{off-res}}$ results from off-resonant AC Stark shifts of components of the laser field with all allowed transitions to off-resonant hyperfine states (not pictured in Fig. \ref{image:Lambda-scheme}.a) \cite{Vanier_1989, Zhu_PTTI_2000}. For a single frequency component of the light, $m$, the shift is given by
\begin{equation}\label{ACStarkShift}
\Delta_{\text{off-res},m} = \frac{1}{4} \sum_{n} \left[ \frac{\delta_{m,n,2} |\Omega_{m,n,2}|^2}{(\delta_{m,n,2})^2 + (\gamma/2)^2} -\frac{ \delta_{m,n,1} |\Omega_{m,n,1}|^2}{(\delta_{m,n,1})^2 + (\gamma/2)^2}\right].
\end{equation}
To get $\Delta_{\text{off-res}}$, $\Delta_{\text{off-res},m}$ must be calculated for each CPT frequency component $m$ and summed. $\delta_{m,n,1}$ and $\delta_{m,n,2}$ are the one-photon detunings of the light components $m$ from the lower states ($|1\rangle$ and $|2\rangle$ in Fig. \ref{image:Lambda-scheme}.a) to the n$^\text{th}$ off-resonant state, and $\Omega_{m,n,1}$ and $\Omega_{m,n,2}$ are the corresponding Rabi frequencies of the light components $m$. In practice, for $D1$ $^{87}$Rb interrogation, the largest contribution to the shift arises from coupling to the non-resonant exited-state hyperfine level detuned by $\sim 815$ MHz.

A simple analytical expression for the off-resonant shift for one-photon Ramsey spectroscopy of optical transitions in two-level atoms was previously developed  \cite{Yudin_2009,Yudin_2010}. 
There, a Taylor-series expansion of the excited-state population around the central Ramsey fringe was found in terms of $\Delta_{\text{off-res}}$. The main contribution to the shift of the central fringe was shown to be proportional to $T^{-1} (\Delta_{\text{off-res}}/\Omega_0)$, where $\Omega_0$ is the one-photon Rabi frequency during the pulses. 

The nature of CPT Ramsey spectroscopy is fundamentally different than one-photon Ramsey spectroscopy. Three energy levels are involved in the transitions and expressions for the Ramsey fringes are necessarily much more complex. Also, the atomic populations do not undergo Rabi oscillations as they enter the coherent superposition (dark) state, but rather are pumped into that state with a Raman pumping rate $\Gamma_p$ \cite{Hemmer_JOSAB_1989,Zanon_EFTF2004,Guerandel_IEEETM_2007}. $\Gamma_p$ is proportional to the field intensity $I \propto \left|E\right|^2$ instead of the field amplitude typical of one-photon transitions. 

Nevertheless, we can show that a expression for the off-resonant shift similar to the one developed for one-photon transitions \cite{Yudin_2009,Yudin_2010} holds for two-photon CPT transitions by making a simple substitution and comparing the resulting expression to numerical solutions to the density matrix equations. Because of the two-photon nature of the CPT transitions, we substitute the Rabi frequency with the rate for pumping atoms into the dark state, $\Gamma_p$, and we represent the shift (in Hz units) for CPT Ramsey spectroscopy as
\begin{equation}\label{shift_Ramsey_general}
    \bar{\delta}_\text{CPT-R} = \frac{A}{2\pi T}\cdot\frac{\Delta_{\text{off-res}}}{\Gamma_p}, \qquad
    \Gamma_p = \frac{1}{4}\frac{|\Omega_{1}|^2+|\Omega_{2}|^2}{\gamma_\text{opt}}.
\end{equation}
Here $A$ is a coefficient whose dependence on experimental parameters we determine below through numerical calculations of shifts in the central Ramsey fringe performed with the density matrix equations and experimental studies.

\begin{figure}[t]
		    \includegraphics[width=7.0cm]{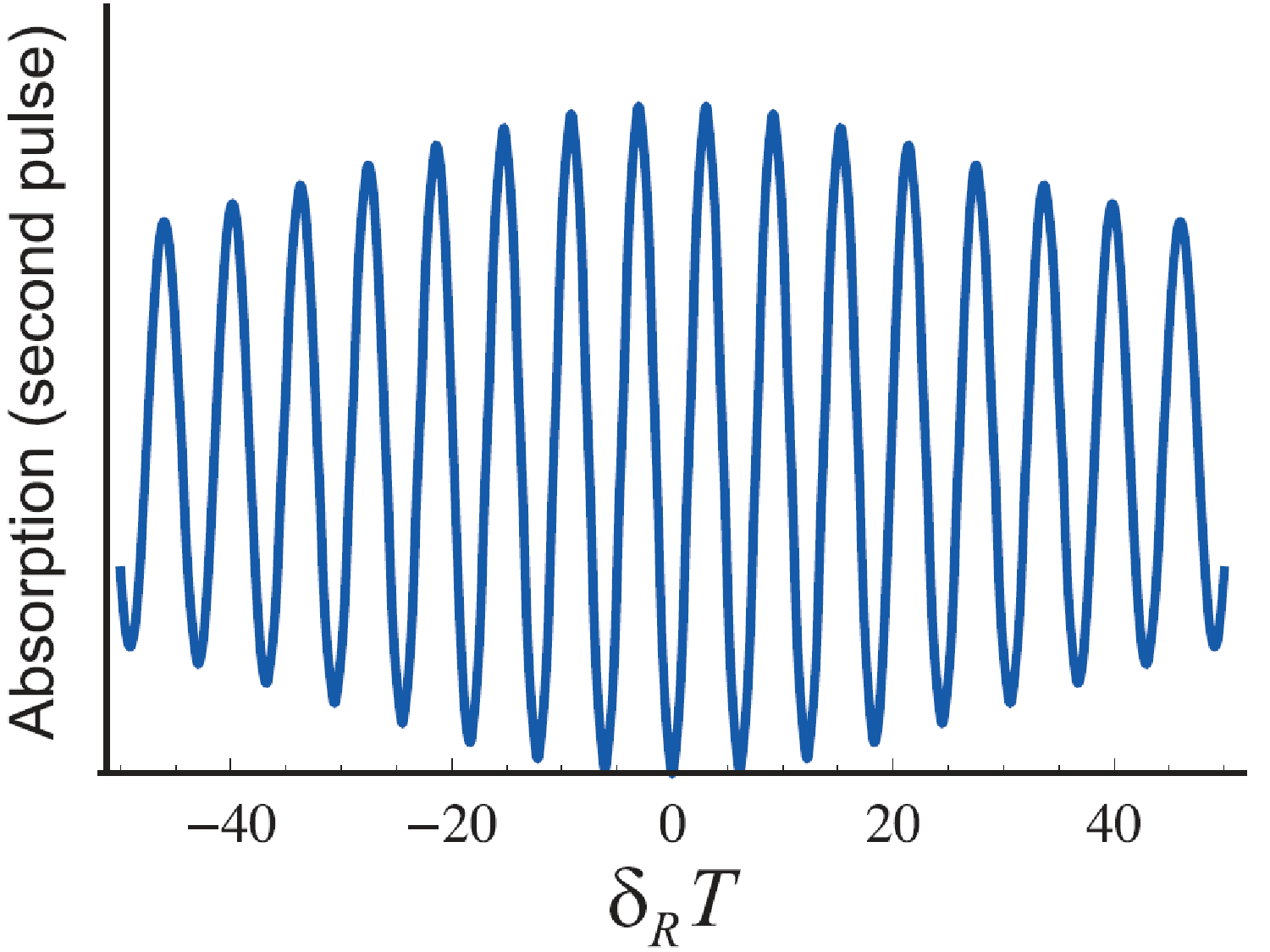}
		\caption{
		\label{Spectra}
   Typical Ramsey fringes simulated with Eqs. \ref{rho_eqs} under condition of $\Gamma_p T \gg 1$.  }
\end{figure}

To calculate the fringes, we use the absorption (spontaneous scattering), which is proportional to $\int_{t}^{t+\tau_2}\rho_{33}(t')dt'$ integrated over the second pulse $\tau_2$ ( Fig. \ref{image:Lambda-scheme}). A typical spectrum of the Ramsey fringes is shown in Fig. \ref{Spectra} under the condition where $\Gamma_{\text{p}}T\gg 1$, where the width (FWHM) of the narrow individual fringe is equal to the value $1/2T$ (in Hz units), and the wide envelope has the width of order of $2 \Gamma_p$. In our experiments, we typically work in the regime where $\Gamma_{\text{p}}T > 400$ and several hundred fringes are visible.

\begin{figure}[t]
		    \includegraphics[width=8.0cm]{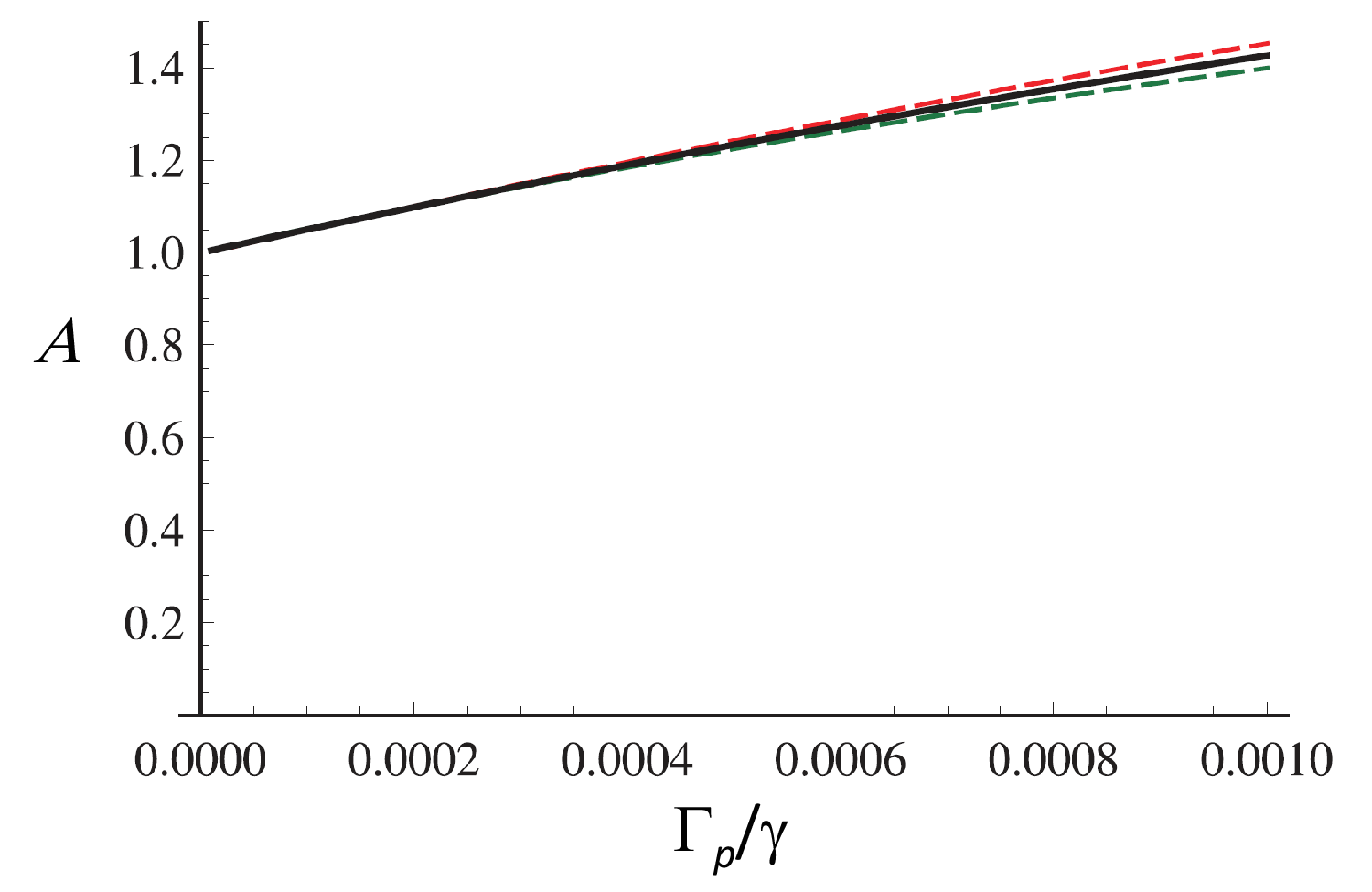}
		\caption{
		\label{ASims}
   Dependence of $A$ on $\Omega_{\text{eff}}$: $|\Omega_1|=|\Omega_2|$ and $\gamma_1=\gamma_2$ (solid black line); $|\Omega_1/\Omega_2|=4$ and $\gamma_1/\gamma_2=0.25$ (dashed red line);  $|\Omega_1/\Omega_2|=0.25$ and $\gamma_1/\gamma_2=0.25$ (dashed green line). Calculations are done with the following parameters: $\gamma_\text{opt}=\gamma/2$, $\Gamma_0=0$, $\delta_\text{1-ph}=0$, $\tau_1=\infty$, $\tau_2=10^3\gamma^{-1}$, $T=10^6 \gamma^{-1}$.
    }
\end{figure}

To determine which parameters $A$ depends on, we performed numerical simulations with Eqs. \ref{rho_eqs} using parameters typical to our cold-atom CPT experiments: $\delta_\text{1-ph}=0$, $\gamma_\text{opt}=\gamma/2$ (i.e., pure spontaneous relaxation, no buffer gas broadening), and $\Gamma_0=0$ (due to the absence of atom-atom collisions). We assume the steady state solution for Eq. (\ref{rho_eqs}) is realized during the first Ramsey pulse, and take $A$ to be independent of $\tau_1$. Finally, we stay in the small saturation regime ($\Gamma_p/\gamma\ll 1$) where the duration of the second pulse, $\tau_2$, is less than the typical time to pump the atoms into the dark state.

For these conditions, numerical simulations show that for $|\Delta_{\text{off-res}}/\Gamma_p|\ll 1$, $A$ does not depend on $\Delta_{\text{off-res}}$ ($|\Delta_{\text{off-res}}/\Gamma_p|< 0.01$ is typical  for our experiments), and, if $\Gamma_pT\gg 1$ and $|\Delta_{\text{off-res}}/\Gamma_p|\ll 1$, $A$ weakly depends on $T$. In Fig. \ref{ASims}, the calculated dependence of $A$ on $\Gamma_{p}$ is presented for different ratios of $|\Omega_1/\Omega_2|$ and $\gamma_1/\gamma_2$. 
As is evident by the curves in Fig. \ref{ASims}, the coefficient $A$ can be approximated by:
\begin{equation}\label{A_univ}
 A\approx 1+0.5\tau_2\Gamma_p.
\end{equation}
It is noteworthy that the derived expression for the off-resonant light shift, Eq. (\ref{shift_Ramsey_general}), demonstrates similar behavior to earlier expressions derived by Yano et al. \cite{Yano_IFCSEFTF_2015} using a completely different approach. In our experiments with $^{87}$Rb, the typical low intensity and short second Ramsey pulse lead to values of $A$ ranging from $1$ to $3$ depending on interrogation scheme. 

In the more general case with other atoms, because the three-level theoretical model does not consider Zeeman substructure of real atoms and other complexities, it is necessary to parameterize $A$ as 
\begin{equation}\label{A_general}
A = A_0 + A_1 \tau_2 \Gamma_p,
\end{equation} 
where $A_0$ and $A_1$ depend on the type of atom, the interrogation scheme, the angular momenta of the resonant hyperfine levels, and the polarization configuration of the light beams.

Since both $\Delta_{\text{off-res}}$ and $\Gamma_p$ are proportional to the total intensity, $\Delta_{\text{off-res}}/\Gamma_p$ does not depend on intensity, and the only intensity dependence comes from $A$, parameterized through $A_1$. Thus, Eq. (\ref{shift_Ramsey_general}) predicts significantly reduced sensitivity of $\bar{\delta}_\text{CPT-R}$ to fluctuations of the total intensity as compared to CW interrogation. The main source of the field-induced fluctuations of $\bar{\delta}_\text{CPT-R}$ are fluctuations of the intensity ratio ($R = I_2/I_1$) where $I_1$ and $I_2$ are the intensities of the CPT light fields (Fig. \ref{image:Lambda-scheme}.a).

Recall that for continuous-wave spectroscopy the shift of the clock transition is equal to $\Delta_{\text{off-res}}\propto I$ (Eq. (\ref{ACStarkShift})) and, therefore, this shift varies with fluctuations of the field intensity $I$. Thus, in the context of clock stability, these theoretical calculations show a basic advantage of the CPT Ramsey spectroscopy in comparison with the usual continuous-wave CPT spectroscopy.

\section{Experiments}
\label{Experiments}

To measure the off-resonant light shifts we used the cold-atom CPT clock described previously \cite{Liu_PRApplied_2017}. In this apparatus, a frequency synthesizer is used to modulate the interrogation light at the hyperfine ground-state splitting, and quantum interference within the atoms causes them to stop absorbing light. The hyperfine CPT resonance of the atoms is probed by detecting the transmission of the laser pulse and is used to steer the frequency of the synthesizer, regulating it to the atomic resonance. The sequence of operation of the CPT clock is described hereafter. First, $^{87}$Rb atoms are cooled and trapped in a magneto-optical trap  with a typical cooling period of $20$ ms followed by a $3$ ms molasses period. The atoms are then released and interrogated using a Ramsey spectroscopy sequence (Fig.~\ref{image:Lambda-scheme}.b) while in free fall, with a small magnetic field  applied to set the quantization axis to the direction of the CPT beam. Due to the atoms falling, the total Ramsey sequence (with duration $\tau_1 + T + \tau_2$) is limited to about $20$ ms by the $3.6$ mm ($1/e^2$) diameter of the CPT beam. The overall cycle of cooling and interrogation takes up to $45$ ms.  

The atoms are interrogated with CPT light resonant with the D1 transition at $795$ nm. The two CPT light-fields are generated by driving a fiber-coupled electro-optic phase modulator (EOM) at the $^{87}$Rb hyperfine-splitting frequency ($\approx6.835$ GHz). The off-resonant light shift depends strongly on the intensity ratio. Therefore, the light exiting the EOM is sampled with a Fabry-Perot cavity (FP), and the ratio is measured by curve-fitting the peaks in the FP transmission curve. 

\begin{figure}
	\includegraphics[width=8.6cm]{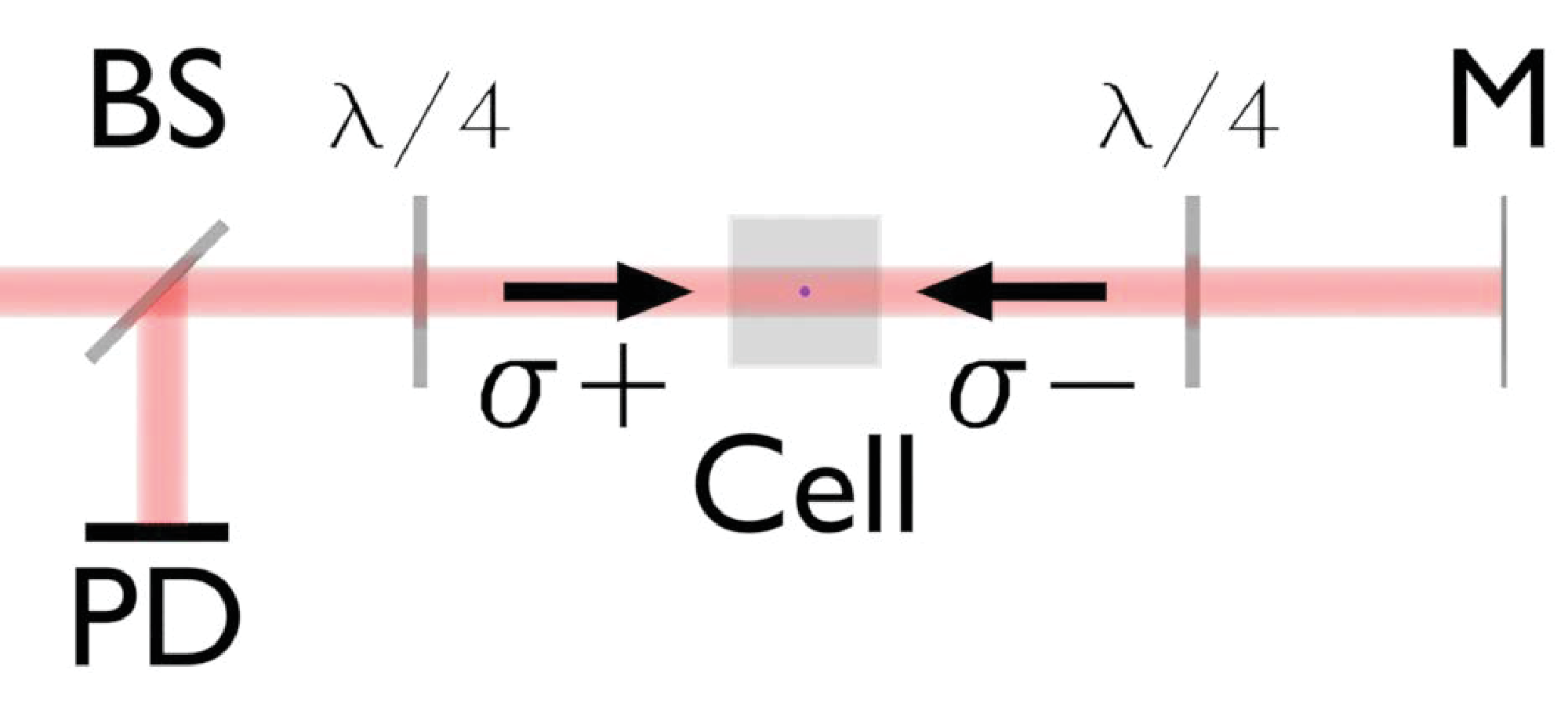}
	\caption{\label{image:Experiment} A simplified diagram of the experimental set-up for the $\sigma_+ - \sigma_-$ interrogation scheme. To realize the lin$||$lin scheme the $\lambda/4$ waveplates are removed and the location of the retro-reflecting mirror (M) is adjusted. (BS: Beam splitter, PD: photodiode, M: retro-reflecting mirror)}
\end{figure}

To obtain high CPT contrast, we use either the lin$||$lin \cite{Taichenachev_JETP_2005,Zibrov_2010} or $\sigma_+ - \sigma_-$ \cite{Taichenachev_2004,Kargapoltsev_LPL_2004} interrogation scheme, both of which probe double-$\Lambda$ systems that prevent the atoms from being trapped in the end magnetic sublevels. The two counter-propagating CPT beams are realized by retro-reflecting the CPT beam and measuring the transmitted power after the second pass through the atoms (Fig. \ref{image:Experiment}). For all measurements shown here, the ground-state relaxation was limited by background collisions and had negligible contribution to the linewidth.

The Ramsey pulse sequence consists of two pulses separated by a dark period (Fig. \ref{image:Lambda-scheme}.b). The first pulse with duration $\tau_1$ prepares an atomic coherence, or dark state, between levels $|1\rangle$ and $|2\rangle$, $T$ is the free evolution interval, and the second pulse with duration $\tau_2$ is for detection. The pulse sequence is generated by controlling the RF input to a double-pass acousto-optic modulator. The clock is typically operated with a CPT beam average input intensity of $1.33$ W/m$^2$ at the atoms' location.

\begin{figure}[t]
    \includegraphics[width=8.6cm]{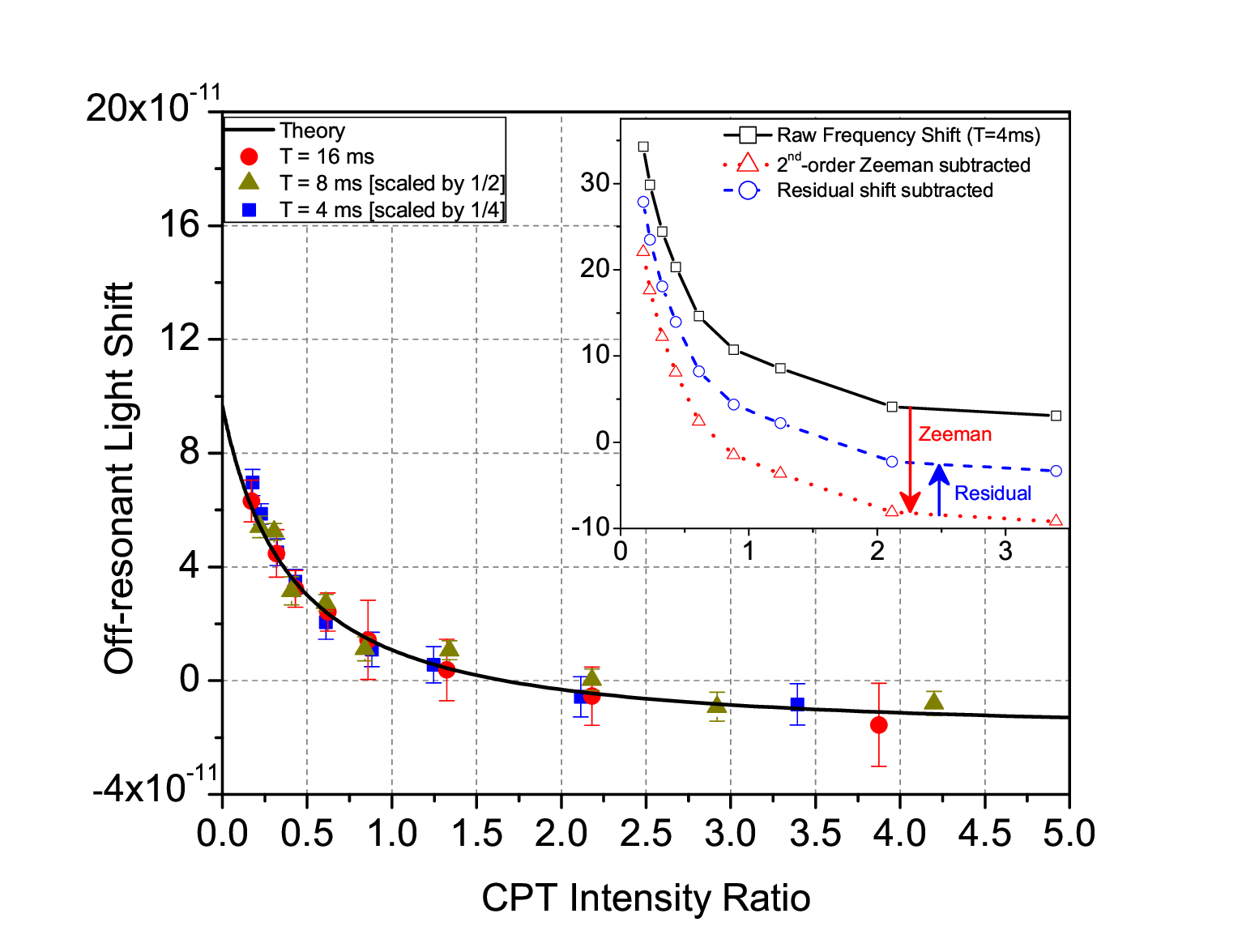}
     \caption{\label{VSTR}
     Comparison between experiment and theory for off-resonant light shifts in fractional frequency units versus intensity ratio. Measurements are for lin$||$lin interrogation with  $T =  4, 8,$ and $16$ ms. The shifts for $T=4$ and $8$ ms are scaled by a factor of $1/4$ and $1/2$, respectively, for easy comparison to the $T=16$ ms measurement and theory. The inset shows the raw measurements for $T=4$ ms, the subtraction of the measured quadratic Zeeman shift ($1.2 \times 10^{-10}$  for our experimental conditions), and the removal of a small residual shift that we attribute to residual Doppler and resonant light shifts in these data. We determined the subtracted residual shift by matching the zero-crossing point of the theory for each data set.
		Before applying the $1/T$ scaling for the 4 and 8 ms data, the residual shifts are $8.8 \times 10^{-12}$ ($T = 16$ ms), $5.9 \times 10^{-11}$ ($T = 8$ ms), and $5.9 \times 10^{-11}$ ($T = 4$ ms). The axis labels for the inset are the same as the main figure. For all measurements, the total intensity was 1.33 W/m$^2$. 
    }
\end{figure}

The absolute frequency shift of the central CPT Ramsey fringe from the accepted value of the hyperfine splitting of $^{87}$Rb is measured by locking a synthesizer to the central Ramsey fringe and comparing the stabilized RF frequency to a hydrogen maser reference. To accurately measure just the off-resonant light shifts, we took steps to mitigate other systematic shifts. The Doppler shift was minimized by applying the CPT beams orthogonal to $\vec{g}$ and retro-reflecting the CPT beam \cite{Esnault_PRA_2013}. The first CPT pulse was made long enough for the system to reach a nearly complete dark-state and minimize resonant light shifts  \cite{Liu_PRApplied_2017}. Finally, the quadratic Zeeman shift was calculated by measuring the quantization magnetic field using a magnetically sensitive atomic transition, and subtracted off. We removed a small residual shift in Fig. \ref{VSTR} from each data set in order to match the zero-crossing point for the theory.

Light shift measurements have been performed versus total CPT intensity,  intensity ratio,  Ramsey period, and  CPT interrogation scheme. The results have been compared to the theoretical model in Eq. (\ref{shift_Ramsey_general}). 

Figure \ref{VSTR} shows off-resonant light shifts versus intensity ratio measured with the lin$||$lin interrogation scheme for three different Ramsey periods $T=4, 8,$ and $16$ ms. The data are presented as absolute frequency shifts in fractional frequency units, by first subtracting the accepted value for the hyperfine ground-state splitting of $^{87}$Rb (6,834,682,610.90 Hz) and then dividing the result by that frequency splitting. The ratio was scanned from $0.25$ to $4$, while keeping $\Gamma_p$ nearly constant. The measurements fit a universal curve (Eqs. \ref{shift_Ramsey_general} and \ref{A_univ}), showing that the shifts are inversely proportional to $T$ and depend strongly on the intensity ratio as predicted.

\begin{figure}[t]
    \includegraphics[width=8.6cm]{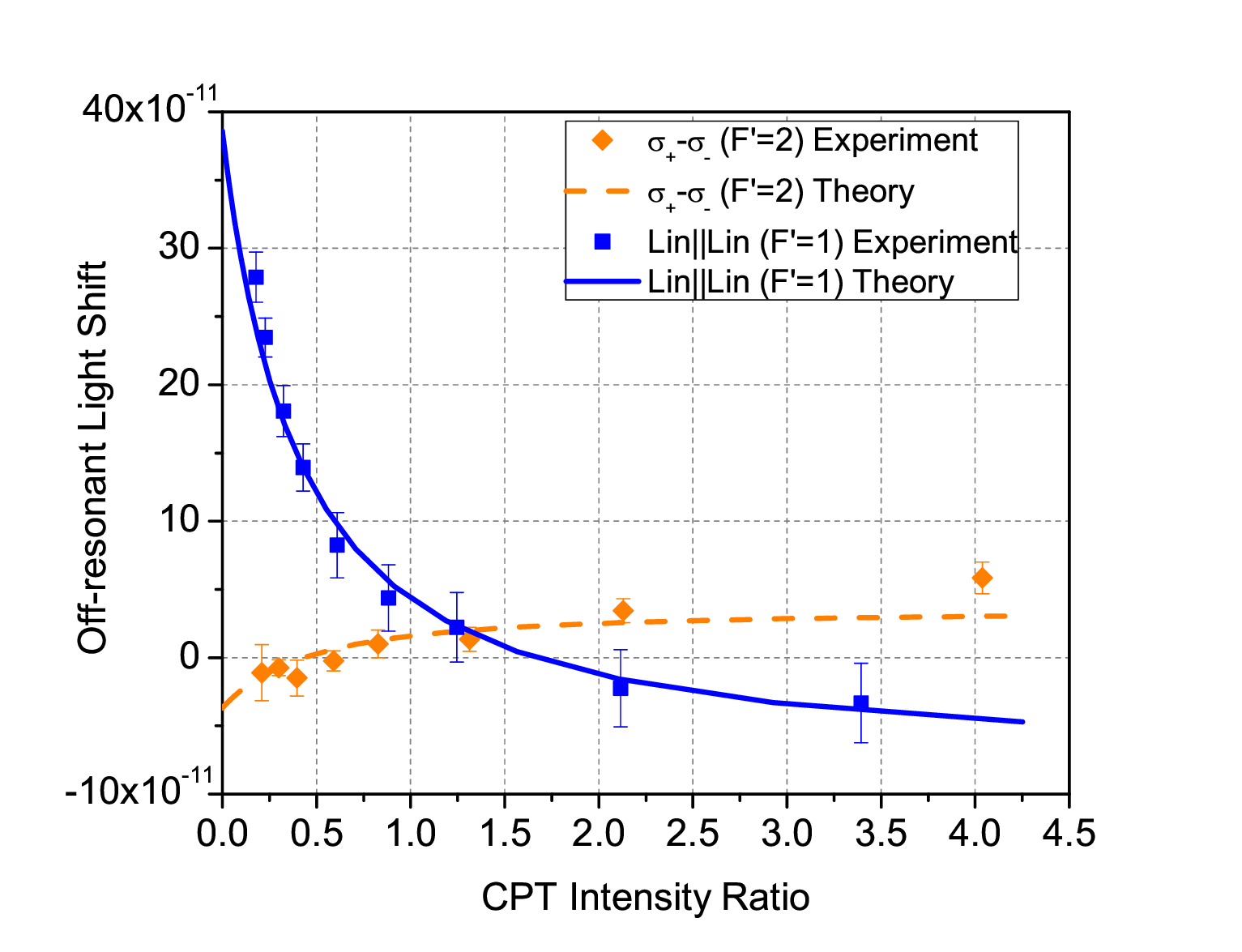}
     \caption{\label{VSpol}
     Comparison between experiment and theory for off-resonant light shifts versus intensity ratio with $T = 4$ ms for the lin $||$ lin (with F'=1) and $\sigma_+ - \sigma_-$ (with F'=2) interrogation schemes. Only the measured Zeeman shift, and not a residual shift, was subtracted from the data. The magnitude of the slope of the theoretical curves at the commonly used intensity ratio of $R=1$ is about $5 \times$ smaller for the $\sigma_+ - \sigma_-$ curve than it is for the lin $||$ lin curve. For all measurements, the total intensity was 1.33 W/m$^2$. We note that the $\sigma_+ - \sigma_-$ configuration is more sensitive to optical alignment, and special care should be taken to ensure proper alignment of the reflected CPT beam (e.g. using a cat-eye setup).}
\end{figure}

Figure \ref{VSpol} shows the off-resonant light shifts versus intensity ratio, with $T=4$ ms, for the two interrogation schemes: lin$||$lin and $\sigma_+ - \sigma_-$ resonant with the $F'=1$ and $F'=2$ levels, respectively. Overall, the  $\sigma_+ - \sigma_-$ off-resonant shifts are significantly smaller  and less sensitive to intensity ratio fluctuations (the curve is flatter) when compared to the shifts for the lin$||$lin scheme. Thus, $\sigma_+ - \sigma_-$ shows an advantage over lin $||$ lin in this atomic system. This advantage originates from the dipole matrix elements of the specific optical transitions that contribute to $\Delta_{\text{off-res}}$ and $\Gamma_p$. For $\sigma_+ - \sigma_-$ interrogation locked to $F'=2$, $\Delta_{\text{off-res}}$ is smaller while $\Gamma_p$ is larger, leading to overall smaller shifts (Eq. (\ref{ACStarkShift})). 

\begin{figure}[t]
    \includegraphics[width=8.6cm]{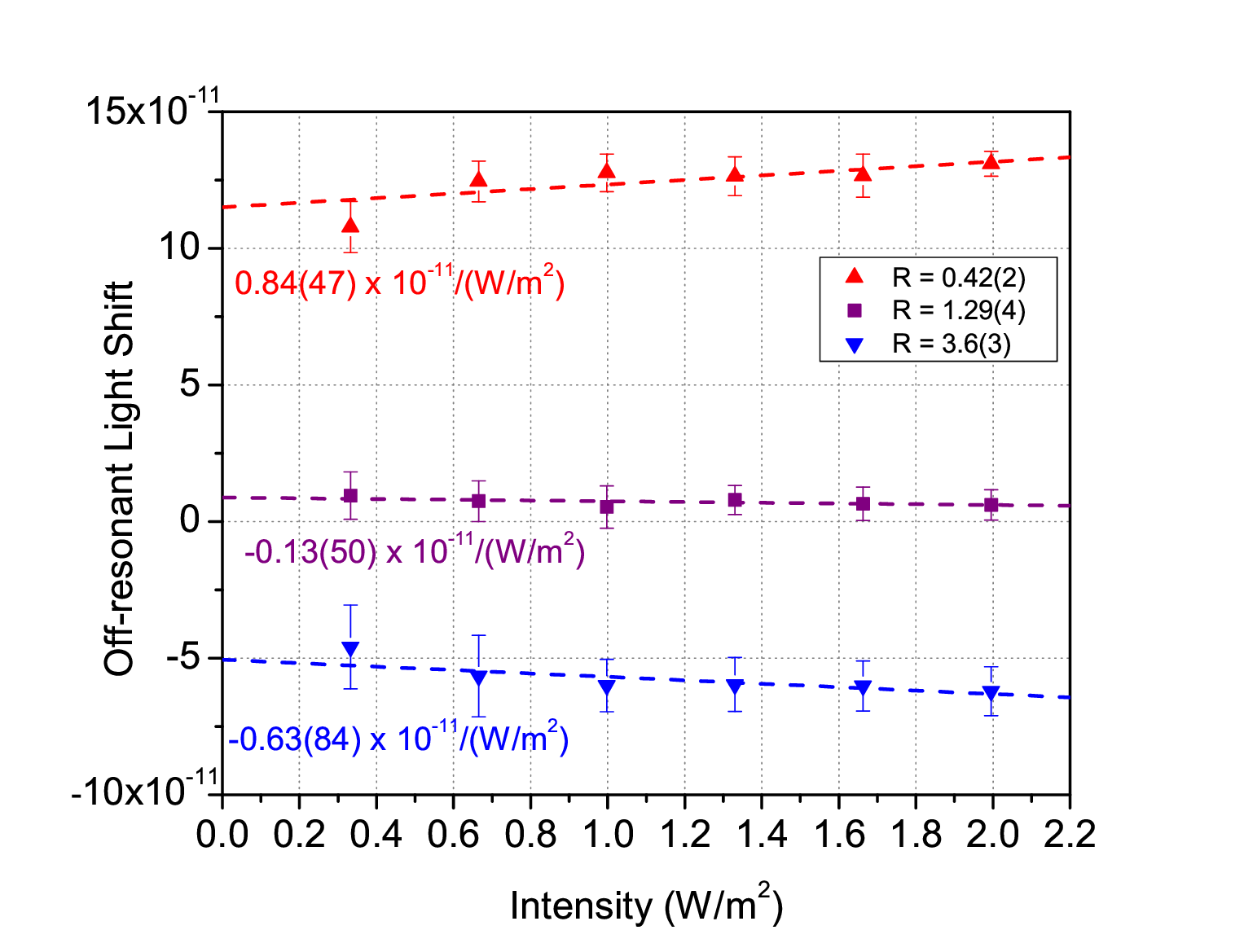}
     \caption{\label{VSI}
     Off-resonant light shift versus total CPT intensity for the lin$||$lin scheme with $T=4$ ms, and three different intensity ratios. Weighted linear fits to the intensity dependencies are shown as dashed lines, and can be used to estimate the sensitivity to intensity variations. Only the measured Zeeman shift, and not a residual shift, was subtracted.
    }
\end{figure}

Figure \ref{VSI} shows the off-resonant light shifts versus total intensity for lin$||$lin interrogation with $T = 4$ ms, and for three different intensity ratios. For all intensity ratios, the frequency shifts depend weakly on the total intensity. The slope for $R = 1.29$ is consistent with zero (in good agreement with Ref. \cite{Zibrov_2010}), while for $R=0.4$ and $3.6$ the slopes are small but non-zero. For typical clock operation with $T = 16$ ms and $\sigma_+ - \sigma_-$ interrogation with $F' = 2$, the shift intensity dependence would be much smaller due to the $1/T$ dependence of the shift (demonstrated by the data in Fig. \ref{VSTR}) and the smaller shifts for the $\sigma_+ -\sigma_-$ scheme (demonstrated by the data in Fig. \ref{VSpol}).

While the expression for $A$ given in Eq. \ref{A_univ} fits the data in Figs. \ref{VSTR} and \ref{VSpol}, it predicts a stronger intensity dependence than we see in Fig. \ref{VSI}. To account for this, we used the general equation for $A$ (Eq. (\ref{A_general})) and determined $A_0$ and $A_1$ through fitting: $A_0 = 2.1 \pm 0.2$ and $A_1 = 0.06 \pm 0.07$.  The smaller value for $A_1$ (compared with the $0.5$ coefficient in Eq. (\ref{A_univ})) reflects the weaker intensity dependence than what is predicted by the three-level model (Eq. (\ref{shift_Ramsey_general})).

\section{Discussion and Outlook}
\label{Discussion}
In this work we have studied, theoretically and experimentally, the off-resonant light shifts in CPT Ramsey spectroscopy of laser cooled $^{87}$Rb atoms. We derived a simple expression for the shift and demonstrated excellent agreement with various experimental measurements. 

We found that the off-resonant light shift depends strongly on the intensity ratio of the two light fields used in the CPT interrogation, therefore the clock stability will depend on the ratio fluctuations. For both interrogation methods studied, lin$||$lin and $\sigma_+ - \sigma_-$, by properly selecting the CPT intensity ratio, one can choose to minimize either the off-resonant light shift or its dependence on the CPT ratio. Generally, the off-resonant light shifts in the $\sigma_+ - \sigma_-$ $(F'=2)$ scheme are smaller and less sensitive when compared with the  lin$||$lin $F'=1$ scheme (due to the dipole matrix elements associated with the excited state).

When measured at the same intensity, CPT Ramsey spectroscopy reduces the CW light shift by a factor of $\approx T \Gamma_p$. For our experimental conditions, with a typical total intensity of around $1 \text{ W/m}^2$, $T \Gamma_p$ is on the order of $10^3$ or $10^4$. A CPT clock interrogated continuously and operated at total optical intensity $10^3$ or $10^4$ times lower would achieve similarly small light shifts. However, in the context of the clock stability, CPT Ramsey spectroscopy has a basic advantage over continuous-wave CPT spectroscopy in substantially reducing the sensitivity of the off-resonant shifts to intensity fluctuations. 

Small intensity dependences have also been measured with CPT Ramsey spectroscopy in vapor cells based on sodium \cite{Yoshida_PRA_2013} and cesium atoms \cite{Kozlova_IEEE_2014,Yano_APExp_2015}. The theory (Eq. (\ref{shift_Ramsey_general})) developed here may also apply to these systems, in spite of their much faster relaxation rates.

We observed an intensity dependence of the shifts (Fig. \ref{VSI}) at least $4$ times smaller than predicted by Eq. (\ref{A_univ}). In a supplementary measurement we found that the dark-state pumping rate, $\Gamma_p$, is 3-4 slower than predicted by Eq. (\ref{shift_Ramsey_general}), which might explain this discrepancy, see Eq. (\ref{A_univ}).  The slower pumping rate is a matter of further study, but we believe it is in part due to the atoms' complex level scheme compared with the simplified three-level atomic model used to determine Eq. (\ref{A_univ}).

\begin{acknowledgments}

The authors acknowledge J. Elgin, R. Boudot, and C. Oates for technical help and discussions. The Russian team was supported by the Russian Science Foundation (Project No. 16-12-10147). M. Yu. B. was also supported by the Ministry of Education and Science of the Russian Federation (Project No. 3.1326.2017/4.6), and Russian Foundation for Basic Research (projects No.17-02-00570 and No.16-32-60050$\underline{\,\,}$a$\underline{\,\,}$dk). This work is a contribution of NIST, an agency of the U.S. government, and is not subject to copyright. 
\end{acknowledgments}

\bibliography{citations}
\bibliographystyle{apsrev4-1}

\end{document}